\documentclass[12pt]{article}
\usepackage{latexsym} \usepackage{epsf}
\usepackage{cite}
\usepackage{a4}
\usepackage{amsfonts}

\newcommand{\oB}{|_{\partial M}}

\begin{document}
\title{\bf Discrete symmetries of functional determinants}

\author{	  D.V. Vassilevich $^{\rm a}$\thanks{On leave from V.A. Fock
		Department of Theoretical Physics,
     		St. Petersburg University,
     		198904 St. Petersburg, Russia.
     		e-mail: Dmitri.Vassilevich@itp.uni-leipzig.de} 
     ~and~ 
     		A. Zelnikov $^{\rm b}$\thanks{On leave from 
		P.N. Lebedev Physics Institute,
		Leninskii prospect 53, Moscow 117 924, Russia. e-mail:
		zelnikov@phys.ualberta.ca}  \\ {}\\
     {$^{\rm a}$\small\it University of Leipzig, Institute 
     for Theoretical Physics}\\
{\small \it  Augustusplatz 10/11, 04109 Leipzig, Germany}\\ 
{$^{\rm b}$\small\it Theoretical 
Physics Insitute, University of Alberta,}\\
{\small \it Edmonton, Alberta, Canada T6G 2J1}}
\maketitle

\begin{abstract}
We study discrete (duality) symmetries of functional determinants.
An exact transformation of the effective action under the inversion 
of background fields $\beta (x) \to \beta^{-1}(x)$ is found. We show 
that in many cases this inversion does not change functional 
determinants. Explicitly studied models include a matrix theory in 
two dimensions, the dilaton-Maxwell theory in four dimensions on 
manifolds without a boundary, and a two-dimensional dilaton theory 
on manifolds with boundaries. Our results provide an exact relation 
between strong and weak coupling regimes with possible applications 
to string theory, black hole physics and dimensionally reduced models.
\end{abstract}
\section{Introduction}
There is an enormous amount of literature devoted to
calculations of the one-loop effective action on various
backgrounds. It is well known that the divergent part of the
effective action is local and can be calculated exactly, at
least in principle. The finite part of the effective action
is typically non-local and usually cannot be obtained in
a closed form. The famous exceptions when the effective
action is known exactly are the Polyakov action \cite{Polyakov}
and the WZNW action \cite{Witten84,Novikov82}. There is also an example
of a spinorial action \cite{Kummer:1999dc} for which 
the finite part of the effective action can be 
calculated exactly. 

In some cases one can obtain powerful relations between
functional determinants and spectral functions of relevant
operators even though the functional determinants (effective
actions) themselves are not known exactly. 
These cases are governed by the Index Theorem (see 
\cite{Gilkey95,EGH} for a detailed review).
The idea behind the Index Theorem is quite
simple. Let the two operators $\Delta_1$ and $\Delta_2$
be represented in a factorized form: $\Delta_1=Q_1Q_2$,
$\Delta_2=Q_2Q_1$. In such a case, the eigenvalues of 
$\Delta_1$ and $\Delta_2$ coincide up to zero modes.
Consequently, differences of all spectral functions
of $\Delta_1$ and $\Delta_2$ (e.g. the zeta function or
the heat kernel) come from  the zero mode contributions
and typically can be expressed in terms of topological invariants.

In this paper we study relations between the functional
determinants of
\begin{equation}
\Delta_+=-(\beta^{-1}\partial_\mu \beta )
(\beta \partial_\mu \beta^{-1})\quad {\rm and}\quad
\Delta_-=-(\beta \partial_\mu \beta^{-1})
(\beta^{-1}\partial_\mu \beta )\,, \label{Del+-}
\end{equation}
where $\beta (x)$ is a matrix valued function or a
positive scalar field, $\beta =e^{-\phi}$. Below we
demonstrate that in certain cases $\log \det \Delta_+-
\log \det \Delta_-$ can be calculated exactly. In
most cases this difference is just zero.
The two operators are related by the transformation 
$~\beta \to \beta^{-1}$.
Hence, the relation between the determinants that has been
advertised above is a kind of duality relation between
effective actions in strong and weak coupling limits.
The key difference between the problem we study here and  
that of the standard Index Theorem is
the summation over the index $\mu$ in operators (\ref{Del+-}).
Because of this summation the eigenvalues of the operators 
$\Delta_+$ and $\Delta_-$ generically are different. 
Nevertheless, we find a relation between
the functional determinants of these operators even though
this property cannot be extended to other
spectral functions (such as higher heat kernel
coefficients). 

At first sight the operators (\ref{Del+-}) look a little bit
artificial, but in fact they are quite general and 
appear in a number of physical applications. 
Consider, e.g., a (spherical) reduction of a field theory
from higher dimensions (see \cite{Kummer:1999zy} for a
recent review). Let $\psi$ be a minimal scalar field living on 
a $D$-dimensional manifold that has the structure of a 
semi-direct product:
\begin{equation}
(ds)^2=g_{\mu\nu}(x)dx^\mu dx^\nu + e^{-\frac 4{D-2} \phi (x)}
(d\Omega )^2 \,.\label{line}
\end{equation}
Here $g_{\mu\nu}(x)$ is the two-dimensional metric and
$(d\Omega )^2$ is the line element on a $(D-2)$-dimensional
manifold. Both the dilaton $\phi(x)$ and the metric $g_{\mu\nu}(x)$
are assumed to depend only on the first two coordinates $x^{\mu}$. 
Then the action for the scalar field reduces to the 2-dimensional 
action,
\begin{equation}
\int d^Dx\sqrt{g^D} g^D_{ik}(\partial^i\psi )(\partial^k\psi)
\ \to \ \int d^2x \sqrt{g} g^{\mu\nu} e^{-2\phi}
 (\partial_\mu\psi) (\partial_\nu \psi )\,.
\label{redact}
\end{equation}
To quantize the theory one should also fix an inner product
of the quantum fields. 
The reduced inner product may also depend on the dilaton
field:
\begin{equation}
<\psi ,\psi' >=\int d^2x \sqrt{g}e^{-2\varphi(\phi)}
\psi (x) \psi'(x) \,. \label{redinpr}
\end{equation} 
If we start with the natural choice of the measure $\sqrt{g^D}$
for the inner product in $D$ dimensions, then the 
preferable choice for the scalar function $\varphi (\phi )$ in $2D$
is, of course, $\varphi (\phi )=\phi$. For simplicity  
we have omitted in Eqs.(\ref{redact}-\ref{redinpr}) 
an unimportant constant factor of
the volume of the $(D-2)$-dimensional manifold.
By performing the 
functional integration over $\psi$ with the measure defined
by (\ref{redinpr}) we obtain the effective action
\begin{equation}
W[\varphi -\rho ,\phi ]
=\frac 12 \log \det (-e^{\varphi -\rho }
\partial_\mu e^{-2\phi} \partial_\mu e^{\varphi -\rho } )\,,
\label{redeff}
\end{equation}
where the conformal gauge for the two-dimensional metric 
$~g_{\mu\nu}=e^{2\rho}\delta_{\mu\nu}~$ 
has been used. The indices
in (\ref{redeff}) are contracted with the Kroneker symbol.
The dependence of the effective action (\ref{redeff}) on 
$\varphi -\rho$ is governed by the scale anomaly which
can be easily calculated for arbitrary $\varphi -\rho$.
The corresponding effective action can be
obtained by integration of the anomaly \cite{Kummer:1997jr,Mukhanov:1994ax}.
This means, in particular, that 
\begin{equation}
W[\varphi -\rho ,\phi ]=W[\phi ,\phi ]+W^{\mbox{\scriptsize scale}}\,.
\label{scaling}
\end{equation}
where $W^{\mbox{\scriptsize scale}}$ can be easily calculated
by the methods of 
\cite{Kummer:1997jr,Alvarez:1983zi,Dowker:1990ue,Tseytlin:2000jr}.
An explicit expression will be given below (see eq. (\ref{1620})).
Thus, the problem of the calculation of the
effective action (\ref{redeff}) is  reduced to the evaluation of 
the determinant of $\Delta_+$ with $\beta =e^{-\phi}$.

One of the most interesting applications of the reduced
theories is the evaluation of the $s$-wave Hawking radiation
\cite{Mukhanov:1994ax} (see \cite{Kummer:1999zy} for more references).
Typically, $\phi\to -\infty$ corresponds to the asymptotically
flat region and $\phi\to +\infty$ to the region near the
black hole singularity. Thus the transformation $\phi \to -\phi$
indeed maps the strong coupling region to the weak coupling one.
Of course, reduction and quantization do not commute. The difference
is the so called dimensional-reduction anomaly 
\cite{Frolov:2000an,Sutton:2000gm,Cognola:2000xp} 
(see also \cite{Nojiri:1999br}).
The reduction of other models (as, e.g., that of the Maxwell
theory in $D$-dimensions to the dilaton-Maxwell theory in
4 dimensions) can be considered along the same lines.

In fact, the operator $\Delta_+$ with $\beta =e^{-\phi}$
is a representation of the scalar Laplacian with a
potential $V_+=-(\partial^2 \phi )+(\partial \phi )^2$.
Indeed, we have traded one scalar function $V$ for another
scalar function $\phi$.
In a recent paper \cite{GusevZelnikov99}
it has been noted that 3rd order terms in $\phi$ are surprisingly 
absent in the expansion of the effective action in powers of $\phi$. 
In the present paper we explain
this result and extend it to all odd orders of $\phi$.

Another important model where the operators (\ref{Del+-})
appear is the bosonic string.
Consider a model with the action 
\begin{equation}
S=\int d^2x \, G_{ab}(x) \partial_\mu \xi^a \partial^\mu \xi^b \,.
\label{string}
\end{equation}
The natural inner product for the field $\xi^a$ reads:
\begin{equation}
<\xi_{1},\xi_2>=\int d^2 x\, G_{ab}\xi^a_1\xi^b_2 \label{innpr}\,.
\end{equation}
Performing the path integral over $\xi$ with the measure
defined by (\ref{innpr}) we arrive at $\det \Delta_+$ 
with $(\beta^2)_{ab}=G_{ab}$. 
Note that if one expands the ordinary bosonic string
action in small fluctuations of the string coordinates
one more term proportional to the target space Riemann tensor
appears in (\ref{string}). Therefore, (\ref{string}) describes
the one-loop partition function for the bosonic string only
in the case of a flat background metric $G_{ab}$ which is perhaps
not of particular interest. Our analysis is rather motivated
by general ideas of string duality 
(see \cite{Giveon:1994fu,Alvarez:1995dn} for introductory reviews).
The target space duality transformation changes the size of the
target manifold and is used to relate strong and weak
coupling limits (usually with respect to small and large
compactification radii). We go further by considering 
arbitrary matrix transformations with arbitrary dependence
on the world-sheet coordinates, and still find a duality
symmetry in the quantum determinants! Although we cannot
give immediately an exact relation to known formulations of the
$T$-duality, we note that
the one-component case is just the dilaton-shift
problem in the $T$-dual models \cite{SchwarzZeitlin93}.
Also, our transformations interchange Dirichlet and
Neumann boundary conditions (see sec. 3), just as it
happens in the open string duality
\cite{Horava:1989ga,DLP89,Green:1991et,Bianchi:1992eu,
Dorn:1996an,Forste:1996hy}.

Throughout this paper we use zeta function regularization
\cite{DowkerCr76,Hawking77}. In the next section we demonstrate
that functional determinants for the operators (\ref{Del+-})
coincide on a two-dimensional manifold without boundary. Our 
method is similar to that used in \cite{SchwarzZeitlin93}
for the one-component case. We also clarify some subtle
mathematical points, such as non-standard logarithmic
terms in the heat kernel expansion and non-locality of some
of the heat kernel asymptotics. In Sec. 3 we extend
our formalism to manifolds with boundaries. For simplicity,
the one-component case is considered. This time the difference of the
effective actions is non-zero. It is given by a
surface integral plus a non-local contribution of the zero
modes. In Sec. 4 we go to four dimensions and demonstrate that the
the effective action in the dilaton-Maxwell
theory is independent of the sign in front of the background dilaton field.
In that section we extensively use the technique of the
multidimensional Darboux transform \cite{ABI}.

\section{Determinants of  matrix operators in 2D }\label{2Dsec}
Consider the first order differential operators
\begin{eqnarray}
&& A_\mu =
\partial_\mu +\Phi_{\mu}=\beta^{-1} \partial_\mu \beta \,, \nonumber \\
&&B_\mu =-A_\mu^\dag =\partial_\mu -\Phi_{\mu}^\dag =\beta \partial_\mu
\beta^{-1} \,,
\label{defAB}
\end{eqnarray}
where $\beta$ is an hermitian matrix-valued field, $\beta^\dag =\beta$.
$\Phi_{\mu}$ is expressed in terms of $\beta$ :
\begin{equation}
\Phi_\mu =\beta^{-1}(\partial_\mu \beta )\,.
\label{phiPhi}
\end{equation}
From the two first-order operators (\ref{defAB}) we construct 
two ``Laplacians''
\begin{equation}
\Delta_+=- A_\mu B_\mu\,, \qquad\qquad \Delta_-=-B_\mu A_\mu \,.
\label{defD+-}
\end{equation}
Note that $\Delta_+$ is mapped to $\Delta_-$ under $\beta\to\beta^{-1}$.

Our aim is to compare functional determinants (one-loop effective actions)
for the operators $\Delta_+$ and $\Delta_-$. For a general elliptic
operator
the regularized effective action reads
\begin{equation}
W^{\mbox{\scriptsize reg}}=\frac 12 \log \det (D)_{\mbox{\scriptsize reg}}
= -\frac{\mu^{2s}}{2}
\int_0^\infty dt~t^{s-1}
{\rm Tr} (\exp (-tD)) \,, \label{effact}
\end{equation}
where we have introduced the (zeta-function) regularization
parameter $s$, which should be set to zero after calculations,
and a massive parameter $\mu$ needed to make the action
dimensionless. To proceed further we need a definition of
the $\zeta$-function of the operator $D$:
\begin{equation}
\zeta^D (s)= {\rm Tr}(D^{-s})=
\frac 1{\Gamma (s)} \int_0^\infty dt~t^{s-1}
{\rm Tr} (\exp (-tD)) \,. \label{defzD}
\end{equation}

The equations (\ref{effact}) and (\ref{defzD}) are valid for
general elliptic (pseudo) differential operators of positive
order. From now on we use the fact that $D$ is an operator of
Laplace type, i.e. $D$ is an elliptic second-order differential
operator with a scalar leading symbol. For such operators the
$\zeta$ function is regular at $s=0$ (see e.g. \cite{Gilkey95}) 
and, therefore, the regularized effective action is
\begin{equation}
W^{\mbox{\scriptsize reg}}=-\frac{\mu^{2s}}{2} {\Gamma (s)}
\zeta^D (s)= -\frac 1{2s} \zeta^D(0) -\frac 12 \zeta^D(0)'
-\log (\mu ) \zeta^D(0) +O(s) \,,
\label{effact2}
\end{equation}
where prime denotes differentiation with respect to $s$.
We have used $\Gamma (s) =1/s -\gamma_E +\dots $ and
have absorbed the Euler constant $\gamma_E$ in a redefinition
of $\mu$. The pole part of (\ref{effact2}) should be
canceled by a counter-term. The renormalized effective
action reads
\begin{equation}
W^{\mbox{\scriptsize ren}}=-\frac 12 \zeta^D(0)'
-\log (\mu ) \zeta^D(0) \,. \label{renea}
\end{equation}
The $\log (\mu )$ term describes the renormalization
ambiguity. By means of the Mellin transformation
one can demonstrate that $\zeta^D(0)=a_1(1,D)$ where
$a_1$ is defined through the heat kernel 
asymptotics 
\begin{equation}
K(f,t,D)={\rm Tr}(f\exp (-tD))
\cong \sum_{n=0}^\infty t^{n-1} a_1 (f,D)
\label{hkD}
\end{equation}
as $t\to +0$ valid for arbitrary smooth matrix-valued
function $f$.
The coefficient $a_1$ is locally computable for operators
of Laplace type. $\zeta^D(0)'$ is non-local and in general
no closed expression for this quantity is available.

Any operator of Laplace type can be represented
in the form
\begin{equation}
D=-(\hat g^{\mu\nu}\nabla_\mu \nabla_\nu +\hat E)
\label{canform}
\end{equation}
with a suitable auxiliary metric $\hat g$,
covariant derivative $\nabla =\partial +\omega$
and potential $\hat E$. For the operator $\Delta_+$
(\ref{defD+-}) we obtain
\begin{eqnarray}
&&\hat g^{\mu\nu}=\delta^{\mu\nu},\qquad \qquad \qquad 
\omega_\mu =\frac 12 \left( \Phi_\mu -\Phi_\mu^\dag \right),
\nonumber \\
&&\hat E =-\frac 12 \partial_\mu 
\left( \Phi_\mu +\Phi_\mu^\dag \right)
-\frac 14 \left( \Phi_\mu +\Phi_\mu^\dag \right)^2 
-\frac 12 [\Phi_\mu ,\Phi_\mu^\dag ] \label{E+}\,.
\end{eqnarray}

The coefficient $a_1$ reads:
\begin{equation}
a_1 (f, D)=\frac 1{24\pi} {\rm tr}
\int d^2x\sqrt{-\hat g} f  
(\hat R+6\hat E)\; .
\label{a1}
\end{equation}
$\hat R$ is the scalar curvature of $\hat g_{\mu\nu}$ and
$\rm tr$ denotes ordinary trace over all matrix indices.

Consider the variation of the zeta function $\zeta^+$ of
the operator $\Delta_+$:
\begin{equation}
\delta \zeta^+(s)= 2s {\rm Tr} \left( (\delta \phi )
A_\mu B^\mu \Delta_+^{-s-1}
-(\delta \phi ) B_\mu \Delta_+^{-s-1} A_\mu \right)\,.
\label{dz+1}
\end{equation}
where we have defined
\begin{equation}
\delta\phi =-\frac 12 ((\delta\beta)\beta^{-1} +
\beta^{-1}(\delta\beta))\,. \label{6dphi}
\end{equation}
In the first term in (\ref{dz+1}) the operators re-combine
in a power of $\Delta_+$, giving the generalized $\zeta$-function:
\begin{equation}
  2s {\rm Tr} \left( (\delta \phi )
A_\mu B^\mu \Delta_+^{-s-1} \right)=
-2s \zeta^+ (\delta \phi ,s) . 
\label{1stterm}
\end{equation}
The contribution of this term to the variation of the effective
action (\ref{renea})
can therefore be treated exactly. Indeed, since $\zeta^+(f,s)$
is regular at $s=0$, we obtain
\begin{equation}
-\frac 12 \partial_s \left[ 2s {\rm Tr} \left( (\delta \phi )
A_\mu B^\mu \Delta_+^{-s-1} \right)\right]_{s=0}=
\zeta^+ (\delta \phi ,0)=a_1(\delta\phi ,\Delta_+ )\,.
\label{1stterm2}
\end{equation}
This reflects a general property of functional determinants.
If the variation of an elliptic operator $D$ has the form
$\delta D=(\delta\sigma_1)D +D(\delta\sigma_2)$ for
some local operators $\delta\sigma_{1,2}$, then the variation of 
the zeta function reads: 
$\delta \zeta^D(s)=-s{\rm Tr}((\delta\sigma_1 +\delta\sigma_2)D^{-s})$.
Consequently, the variation of $\zeta'$ can be expressed
through the heat kernel 
$\delta \zeta^D(0)=-a_1((\delta\sigma_1 +\delta\sigma_2),D)$.

The second term in (\ref{dz+1}) requires more care
because the operators under the trace cannot be
recombined in a power of $\Delta_+$: 
\begin{eqnarray}
{\rm Tr}\left( (\delta \phi )B_\mu \Delta_+^{-s-1} A_\mu \right)
&=&\frac 1{\Gamma (s+1)} \int_0^\infty dt t^{-s}
{\rm Tr}\left((\delta \phi )B_\mu \exp (-t\Delta_+ )A_\mu \right)
\nonumber \\
&=&-\frac 1{\Gamma (s+1)} \int_0^\infty dt t^{-s}
{\rm Tr}\left((\delta \phi ) F \exp (-tF) \right) \nonumber \\
&=&-{\rm Tr}\left( (\delta \phi ) F^{-s}
\right)\,.
\label{ABtoF}
\end{eqnarray}
Here $F$ is a matrix operator having both space-time vector indices
and ``internal'' indices of the matrix field $\beta (x)$: 
\begin{equation}
F_{\mu\nu}=-B_\mu A_\nu \label{defF} \,.
\end{equation}
In eq. (\ref{ABtoF}) the trace is taken over all matrix indices.

It is clear from the construction that the operator $F$ should
be understood as acting on the space ${\cal V}^L$ of $B$-longitudinal
vector fields which can be represented as $v_\mu =B_\mu \psi$,
where $\psi$ is a scalar function. Such fields satisfy the condition
\begin{equation}
B_\mu \epsilon^{\mu\nu}v_\nu =0 \,.\label{cond1}
\end{equation}
The spectrum of the operator $F$ on ${\cal V}^L$ can be
constructed easily. Let $\psi^\lambda$ be a normalized
 eigenfunction of the
operator $\Delta_+$ with a non-zero eigenvalue $\lambda$.
Then the functions 
$v^\lambda_\mu = B_\mu (\Delta_+)^{-1/2} \psi^\lambda$ 
are normalized eigenfunctions of the operator $F$.
Accordingly, we write the off-diagonal heat kernel 
\begin{eqnarray}
K(F,x,y;t)_{\mu\nu}&=&
\sum_{\lambda} v^\lambda_\mu (x) v^\lambda_\nu (y) e^{-t\lambda}
\nonumber \\ 
&=&\int d^2x'\int d^2y' B_\mu^x (\Delta_+ )^{-1/2}_{xx'}
B_\mu^y (\Delta_+ )^{-1/2}_{yy'} K(\Delta_+,x',y';t)\,.
\label{KF}
\end{eqnarray} 
For the traced heat kernel this relation simplifies to
\begin{equation}
{\rm Tr}(\exp (-tF)) ={\rm Tr}(\exp (-t\Delta_+)) \,.
\label{hkFd+}
\end{equation}

The operator $F$ contains a non-local projector on the
space ${\cal V}^L$. Therefore, $F$ is a {\it pseudo}
differential operator rather than a differential one. According to the
general theory \cite{Gilkey95}, the generalized zeta
function for such operator can have a simple pole
at $s=0$:
\begin{equation}
\zeta^F (f,s)=\frac 1s \zeta^F_{\mbox{\scriptsize pole}}(f)
+\zeta^F_0 (f)+O(s) \,.
\label{zF0}
\end{equation}
By substituting (\ref{zF0}) into (\ref{dz+1}) one obtains
\begin{equation}
\delta\zeta^+(0)'=-2\zeta^+ (\delta \phi ,0)
+2 \zeta^F_0 (\delta\phi ) \,. \label{dz+2}
\end{equation}

Next we relate the coefficient $\zeta^F_0 (f)$ to the heat
kernel asymptotics. We invert the obvious relation
\begin{equation}
\zeta^F (f,s)= {\rm Tr}(fF^{-s})=
\frac 1{\Gamma (s)} \int_0^\infty dt\, t^{s-1}
{\rm Tr} (f\exp (-tF))  \label{defzF}
\end{equation}
to show that
\begin{equation}
K(f,t,F)={\rm Tr} (f\exp (-tF))
=\frac 1{2\pi i} \oint ds\, \Gamma (s)\zeta^F (f,s) t^{-s}\,,
\label{invert}
\end{equation}
where the integration contour encircles all poles of the integrand.
The contribution of the pole at $s=0$ reads
\begin{equation}
\zeta^F_0(f) -(\gamma_E +\log t )\zeta^F_{\mbox{\scriptsize pole}}(f)
\,. \label{polecont}
\end{equation}
The Euler constant $\gamma_E$ can be absorbed again
in re-definition of the normalization scale $\mu$ which
is not shown explicitly in (\ref{defzF}) and (\ref{invert}).
Variation (\ref{dz+2}) is again defined by the $t^0$
term in the heat kernel asymptotics.

We see that the heat kernel $K(f,t,F)$ has a quite
unusual $\log$-term in the small $t$ asymptotics.
Even though this term does not contribute to
the variation of the effective action, it is
instructive to calculate its value. Translating
equation (\ref{KF}) from kernels to the operator
notation we obtain:
\begin{eqnarray}
K(f,t,F)&=&{\rm Tr} \left[ f B_\mu (\Delta_+)^{-1/2}
\exp (-t\Delta_+) \left( B_\mu (\Delta_+)^{-1/2}\right)^\dag
\right] \nonumber \\
&=&-{\rm Tr} \left[ A_\mu f B_\mu
\Delta_+^{-1} \exp (-t\Delta_+)\right] \nonumber \\
&=&{\rm Tr} \left[ f\exp (-t\Delta_+) \right]
-{\rm Tr} \left[ f_{,\mu} B_\mu \Delta_+^{-1} 
\exp (-t\Delta_+)\right] \,, \label{KF2}
\end{eqnarray}
where
\begin{equation}
f_{,\mu}=\partial_\mu f + [\Phi_\mu ,f] \,. \label{fmu}
\end{equation}
The first term on the last line in (\ref{KF2}) obviously
does not contain $\log t$ terms. The second term can be
represented as an integral over the proper time:
\begin{equation}
-{\rm Tr} \left[ f_{,\mu} B_\mu \Delta_+^{-1} 
\exp (-t\Delta_+)\right] = -\int_t^\infty ds
{\rm Tr} \left[ f_{,\mu} B_\mu 
\exp (-s\Delta_+)\right] \,,\label{KF3}
\end{equation}
To evaluate (\ref{KF3}) we use the method of \cite{bgv97}.
Consider the operator $\Delta_+^\epsilon =\Delta_+ -\epsilon
f_{,\mu}B_\mu$. Then
\begin{equation}
{\rm Tr} \left[ f_{,\mu} B_\mu 
\exp (-s\Delta_+)\right] = \frac 1s 
\partial_\epsilon  
{\rm Tr} \left[ \exp (-s\Delta_+^\epsilon )\right]\Big|_{\epsilon =0} \,,
\label{KF4}
\end{equation}
The $1/s$ term in the small $s$ asymptotics of (\ref{KF4})
is immediately calculated with the help of (\ref{a1}):
\begin{equation}
-\frac 1s \frac 1{8\pi} \int d^2x 
{\rm tr}\left( f_{,\mu}(\Phi_{\mu}+\Phi_\mu^\dag )\right)\,,
\label{1s}
\end{equation}
This term generates the $\log t$ term in the small-$t$
asymptotics of (\ref{KF2}) and (\ref{KF3}). We conclude that
\begin{equation}
\zeta^F_{\mbox{\scriptsize pole}}(f)= \frac 1{8\pi} \int d^2x 
{\rm tr}\left( f_{,\mu}(\Phi_{\mu}+\Phi_\mu^\dag )\right)\,,
\label{logt}
\end{equation}
Note that this term is local. This supports the quite
general observation that the leading singularity 
($\log t$ in the present case) is usually local , while
the subleading terms ($t^0$) are non-local.

In order to be able to apply standard expressions for the
heat kernel coefficients it is convenient to extend the
operator $F$ to the space ${\cal V}$ of {\it all}
vector fields (which also carry the internal indices). 
The space ${\cal V}$ splits into a
direct sum
\begin{equation}
{\cal V}={\cal V}^L\oplus {\cal V}^T \,,
\label{dirsum}
\end{equation}
where the space ${\cal V}^T$ consists of the vector
fields $u^\mu =\epsilon^{\mu\nu}A_\nu \psi$. It is easy
to check that the decomposition (\ref{dirsum}) is orthogonal,
i.e. $\int d^2x\,u^\mu v_\mu =0$ for $v\in {\cal V}^L$
and $u \in {\cal V}^T$. It is clear that $F$ is zero on
${\cal V}^T$. Define the operator
\begin{equation}
\tilde F_{\mu\nu}=-\epsilon_{\mu\mu'}\epsilon_{\nu\nu'}
A^{\mu'}B^{\nu'} \, \label{defFtil}
\end{equation}
that is zero on ${\cal V}^L$.

It can be demonstrated that the zero modes of the map
$\psi \to v_\mu \in {\cal V}^L$ (resp. 
$\psi \to u_\mu \in {\cal V}^T$) coincide with the zero
modes of $\Delta_+$ (resp. $\Delta_-$) and that the
operators $F$ and $\tilde F$ have no more zero
modes on ${\cal V}^L$ and ${\cal V}^T$ respectively.
Suppose that the underlying manifold is ${\bf R}^2$
and $\beta (x)$ goes to a non-degenerate
constant matrix as $|x|\to\infty$.
In such a case the zero modes $\psi^0_\pm (x) =\beta (x)^{\pm 1}c$
($c$ is a constant vector in the ``internal'' space) of the
operators $\Delta_\pm$ are non-normalizable and, therefore,
can be neglected.
Zero-modes of the decomposition (\ref{dirsum}) (analogs
of harmonic one-forms) are also absent.

The operator $F+\tilde F$ is an elliptic operator
of Laplace type,
\begin{equation}
\tilde F+F = \delta_{\mu\nu}\Delta_+ -\partial_\nu \Phi_\mu^\dag
-\partial_\mu\Phi_\nu -[\Phi_\nu ,\Phi_\mu^\dag ] \,.
\label{F+F}
\end{equation}
The coefficient $a_1$ for this operator can be found
with the help of (\ref{a1}):
\begin{equation}
a_1(f,\tilde F+F) =-\frac 1{8\pi} \int d^2x\,
{\rm tr} (f(\Phi_\mu +\Phi_\mu^\dag )^2)\,.
\label{a1FF}
\end{equation}

For the heat kernel of the operator $\tilde F$ there is
a representation similar to (\ref{KF}),
\begin{eqnarray}
K(\tilde F,x,y;t)_{\mu\nu} 
&=&\int d^2x'\int d^2y' \epsilon_{\mu\mu'}
 A_{\mu'}^x (\Delta_- )^{-1/2}_{xx'} \nonumber \\
&&\times \epsilon_{\nu\nu'}
A_{\mu'}^y (\Delta_- )^{-1/2}_{yy'} K(\Delta_-,x',y';t)\,.
\label{KFtilde}
\end{eqnarray}
From equation (\ref{KFtilde}) follows an important
observation: the trace over vector indices $K(\tilde F,x,y;t)_{\mu\mu}$
is obtained from $K(F,x,y;t)_{\mu\mu}$ if we exchange 
$\beta$ and $\beta^{-1}$. By repeating the calculations
above for the operator $\Delta_-$ we obtain
\begin{equation}
\delta\zeta^-(0)'=2\zeta^- (\delta \phi ,0)
-2 \zeta^{\tilde F}_0 (\delta\phi ) \,. \label{dz-2}
\end{equation}
Now we subtract (\ref{dz-2}) from (\ref{dz+2}) to
demonstrate that
\begin{equation}
\delta \left( \zeta^+(0)'-\zeta^-(0)'\right)=
-2\zeta^+(\delta\phi ,0)-2\zeta^-(\delta\phi ,0)
+2\zeta^{\tilde F+F} (\delta\phi ,0) \,.
\label{dz+-}
\end{equation}
All zeta functions on the right-hand side of (\ref{dz+-})
correspond to elliptic operators. We can replace them
by the heat kernel coefficients:
\begin{equation}
\delta \left( \zeta^+(0)'-\zeta^-(0)'\right)=
2\left(-a_1(\delta\phi ,\Delta_+)-a_1(\delta\phi ,\Delta_-)
+a_1(\delta\phi ,\tilde F+F)\right)=0\,.
\label{1837}
\end{equation}
This equation demonstrates that the difference
$\zeta^+(0)'-\zeta^-(0)'$ does not depend on $\beta$.
Since this difference is zero for $\beta =\beta^{-1}={\bf 1}$,
\begin{equation}
\zeta^+(0)'=\zeta^-(0)'\,,
\label{zetas}
\end{equation}
or
\begin{equation}
W^{\mbox{\scriptsize ren}}[\beta ]=
W^{\mbox{\scriptsize ren}} [\beta^{-1} ] \,.
\label{symWren}
\end{equation}
In a particular (one-component) case this relation has been
obtained in \cite{SchwarzZeitlin93} where the method of
\cite{Schwarz79} was used.
The equation (\ref{zetas}) can be also formulated as a
Generalized Index Theorem:
\begin{equation}
\log \det (-A_\mu B_\mu )=
\log \det (-B_\mu A_\mu ) \,.
\label{GIT}
\end{equation}
Normally, the Index Theorem does not contain summations
under the determinant but contains contributions
of the zero modes which are absent in the topologically
trivial situation considered here.

There is an interesting special case 
for the one-component matrix $\beta =e^{-\phi}$ where
$\phi$ is an external scalar field (dilaton). Let the
dilaton depend on one coordinate only, say $x^1$.
In this case 
\begin{equation}
\Delta_\pm =-\partial_2^2 +H_\pm ,\qquad
H_\pm =Q_\pm Q_\mp \,,{}  \label{sups}
\end{equation}
where $Q_\pm =\pm \partial_1 +(\partial_1\phi )$ is
(up to a multiplier) the supercharge used in supersymmetric
quantum mechanics (for an introduction see \cite{Schwabl}).
Obviously, the operators $Q_\pm$ are intertwining
the Hamiltonians $H_\pm$, leading to pairing of their
non-zero eigenvalues. This pairing is the key ingredient
of the Witten index construction \cite{Witten82} which
governs dynamical supersymmetry breaking. Since in this
particular case the operators $H_\pm$ commute with $\partial_2^2$,
the eigenvalues of $\Delta_\pm$ coincide up to zero modes with those
of $H_\pm$. Consequently, if we neglect the zero mode
contributions, as we have agreed to do in this chapter,
all traced spectral functions of $\Delta_+$ and $\Delta_-$
should coincide, including $\zeta^\pm (s)$ 
and $K(1,t,\Delta_\pm)$, for {\it arbitrary} value
of their arguments ($s$ and $t$ respectively).
Here comes the main difference from our result. 
If the dilaton $\phi$ depends on both coordinates,
the operators $\Delta_-$ and $\Delta_+$ have, in
general, different eigenvalues. This can be confirmed
for instance by calculation of the next heat kernel
coefficient $a_2(1,\Delta_\pm )$ which differs for
$\Delta_+$ and $\Delta_-$.

\section{Manifolds with boundaries}
Let us extend the results of the previous section to manifolds
with boundaries. For simplicity we consider here the one-component
case only, $\beta =e^{-\phi}$. Accordingly,
\begin{eqnarray}
\Delta_+=-A^\mu B_\mu \,,\qquad && A_\mu =
D_\mu -\phi_{,\mu}=e^\phi D_\mu e^{-\phi} \,, \nonumber \\
&&B_\mu =-A_\mu^\dag =D_\mu +\phi_{,\mu}=e^{-\phi} D_\mu
e^\phi \,.
\label{1AB}
\end{eqnarray}
Since we will use a curved coordinate system near the boundary
we have restored upper and lower indices and introduced the
Riemannian covariant derivative $D_{\mu}$.

As we will see in the next section, the structure of the elliptic
complex plays an important role in our construction. Thus,
we could use one of the admissible sets of boundary conditions
for that complex \cite{Gilkey95}. However, we prefer a more
direct (though also more lengthy) way to derive the boundary
operators. Let the operator $\Delta_+$ act on the
fields satisfying Dirichlet boundary conditions 
\begin{equation}
\psi\oB =0. \label{Dirbc}
\end{equation}
Obviously, $\Delta_+\psi\oB =-A^\mu B_\mu \psi\oB =0$.
Therefore, the vector $v_\mu =B_\mu \psi$ (see (\ref{cond1}))
should satisfy
\begin{equation}
A^\mu v_\mu \oB =0\,.
\label{bcv}
\end{equation} 

To proceed further we need some notations. Let $N^\mu$ be the 
inward-pointing unit normal at the boundary. Let $\tau$ be a coordinate
on the boundary. We can extend this coordinate system to a
neighborhood of the boundary keeping orthogonality $~(N^\tau =0)$.
The extrinsic curvature $k$ is defined as $k=-N^\mu \partial_\mu g_{\tau\tau}
=\Gamma^N_{\tau\tau}$. We can simplify the analysis by choosing 
a coordinate system in which
$g_{\tau\tau}=const.$ along the boundary (this relation cannot be
extended inside the manifold). 

In the coordinate frame defined above
\begin{equation}
A^\mu v_\mu \oB = e^\phi \left[ (\partial_N -k)(e^{-\phi}v_N)
+g^{\tau\tau}\partial_\tau (e^{-\phi }v_\tau )\right]\oB \,.\label{1712}
\end{equation}
A local extension of the condition (\ref{bcv}) on the space of
all vector fields ${\cal V}$ (\ref{dirsum}) can be obtained
by requiring that $v_N$ and $v_\tau$ satisfy these boundary
conditions independently:
\begin{equation}
(\partial_N -k)(e^{-\phi}v_N )\oB =0,\qquad\qquad
v_\tau \oB =0\,.
\label{bcv2}
\end{equation}
In particular, the vectors 
\begin{equation}
u_\mu =\sqrt{g}\epsilon_{\mu\nu}A^\nu \chi \in {\cal V}^T
\end{equation}
should satisfy (\ref{bcv2}). This leads to Neumann boundary
conditions for the scalar field $\chi$:
\begin{equation}
\partial_N e^{-\phi} \chi \oB =0 \,. \label{bcchi}
\end{equation}
In our procedure the field $\chi$ forms a functional space
for the operator $\Delta_-$. Therefore, the conditions (\ref{bcchi})
become the boundary conditions for $\Delta_-$. Since (\ref{bcchi})
depends on the dilaton, the variation of $\det \Delta_-$ contains
two contributions, one coming from variation of the operator,
the other from the variation of the functional space on which this
operator acts. At this time we do not know how to evaluate
the second contribution. To avoid difficulties with the 
$\phi$-dependence of (\ref{bcchi}) we impose a Neumann boundary
condition on the dilaton:
\begin{equation}
\partial_N\phi \oB =0 \,.\label{bcphi}
\end{equation}
Due to (\ref{bcphi}) the conditions (\ref{Dirbc}), (\ref{bcv2})
and (\ref{bcchi}) are equivalent to the following
$\phi$-independent set:
\begin{eqnarray}
&&\psi \oB =0 \,, \nonumber \\
&&(\nabla_N-k)v_N\oB =0\,,\qquad v_\tau \oB =0 \,,
\nonumber \\
&&\nabla_N\chi \oB =0 \,,\label{relbc}
\end{eqnarray}
where one can recognize the relative boundary conditions for
the de Rham complex \cite{Gilkey95}.

Let us discuss now the zero mode structure of the theory.
To be more precise let us restrict ourselves to compact manifolds
$M$ with the topology of a two-dimensional disk. The only
zero mode of $\Delta_+$ is eliminated by the Dirichlet boundary
condition. There is a zero mode $\psi_-^0=e^{\phi}$ of
$\Delta_-$. There are no zero modes in the vector sector.
This last statement can be derived from \cite{Gilkey95} or
from the simpler analysis of \cite{Vassilevich:1995we}.

The boundary conditions (\ref{relbc}) are mixed. This means
that one of the components of the vector satisfies a Dirichlet
and the other one a Neumann boundary condition. The heat kernel
expansion for this type of the boundary conditions has been
studied in 
\cite{Moss90,BransonGilkey90,Vassilevich:1995we,Branson:1999jz}.
Let us write the boundary conditions for a multi-component field
$v$ in the form:
\begin{equation}
\left( \Pi_D v +(\nabla_N +{\cal S})\Pi_N v \right) \oB =0
\,,
\label{mbc}
\end{equation}
where $\Pi_{D,N}$ are two complementary projectors and ${\cal S}$
is some function on the boundary which can be matrix-valued.
Clearly, the mixed boundary conditions (\ref{mbc}) contain
Dirichlet and Neumann conditions as particular cases.
The boundary term
\begin{equation}
a_1(f,D)=\frac 1{24\pi}{\rm tr} \int dy \sqrt{h}
\left( f(2k +12{\cal S}\Pi_N ) +3f_{,N}(\Pi_N -\Pi_D)
\right) \label{boua1}
\end{equation}
should be added to the volume part (\ref{a1}). Here $y$ is
a coordinate on the boundary and $h$ is the determinant of the induced
metric. 

Obviously, the contribution of the zero mode $\psi^0_-=e^\phi$ of the
operator $\Delta_-$ to the heat kernel reads
\begin{equation}
K_0(\Delta_-,x,y;t) =\frac{\psi^0_-(x)\psi^0_-(y) }{\int d^2z\sqrt{g}
(\psi^0_-(z))^2 } \,.
\label{zerom}
\end{equation}
There is no $t$-dependence on the right hand side of (\ref{zerom}).
Therefore we can easily derive the zero mode contribution to
$a_1(f,\Delta_-)$:
\begin{equation}
a_1^{(0)}(f,\Delta_-)=\frac{\int d^2 x\sqrt{g} f e^{2\phi}}{\int d^2z
\sqrt{g}e^{2\phi}} \,. \label{a1zero}
\end{equation}
By acting exactly as in the previous section we obtain
\begin{eqnarray}
\frac 12 \left( \delta\zeta^+(0)'-\delta\zeta^-(0)'\right)
&=&-a_1^+(\delta\phi )
-(a_1^-(\delta\phi )-a_1^{(0)}(f,\Delta_-)) +a_1^{F+\tilde F}(\delta\phi ) 
\nonumber \\
&=&-\frac 1{2\pi} \int_{\partial_M} dy\sqrt{h} k\delta\phi +
\frac{\int d^2 x\sqrt{g} (\delta\phi ) e^{2\phi}}{\int d^2z
\sqrt{g}e^{2\phi}}\,.
\end{eqnarray}
This equation can be integrated to give (with the help of (\ref{renea}))
\begin{equation}
W^{\mbox{\scriptsize ren}}(\Delta_+)-
W^{\mbox{\scriptsize ren}}(\Delta_-)=\frac 1{2\pi }
\int_{\partial M} dy\sqrt{h}  k\phi -
\frac 12 \log \left( \int d^2z
\sqrt{g}e^{2\phi} \right) +W^{[0]} \,. \label{dWbou}
\end{equation}
Equation (\ref{dWbou}) represents the main result of this section.
Here $W^{[0]}$ describes the difference between the two effective
actions for $\phi =const.$ Since the boundary conditions for
$\Delta_+$ and $\Delta_-$ are different there is no reason
to believe that this difference is zero. For the standard
Euclidean disk $W^{[0]}$ has been calculated in 
\cite{Bordag:1996zc,Dowker:1996bh}. The non-local term
in (\ref{dWbou}) is typical for the effective action
on a compact manifold \cite{Dowker:1994rt}.

As a consistency check
let us show that if $\phi =\phi_c =const$, the right hand side
of (\ref{dWbou}) does not depend on $\phi_c$. The first term
reads $ \chi\phi_c$ where $\chi$ is the Euler
characteristic. The $\phi_c$ dependent part of the second
term is $-\phi_c$. Since we have assumed that $M$
has the topology of the 2-disk ($\chi =1$)  the two contributions
cancel against each other.

To make our discussion self-contained we conclude this section
with a calculation of the change in the effective action under
a conformal transformation of the external metric. Since any 
metric in two dimensions is conformally trivial this will
provide an extension of our previous results 
to arbitrary curved manifolds. We use the methods of Refs. 
\cite{Kummer:1997jr,Alvarez:1983zi,Dowker:1990ue,Tseytlin:2000jr}.
Consider the operator $\Delta^{[\hat \rho]}_+=e^{-2\hat\rho}\Delta_+$,
which  is unitarily equivalent to the operator (\ref{redeff})
$\hat \rho =\rho -\varphi +\phi$. It is clear that under variation
of $\hat \rho$ the effective action changes as
\begin{equation}
\delta W^{\mbox{\scriptsize ren}}(\Delta^{[\hat \rho]}_+)=
-a_1(\delta\hat\rho ,\Delta^{[\hat \rho]}_+) \,.
\label{varWrho}
\end{equation}
The coefficient $a_1$ on the right hand side of (\ref{varWrho})
is given by equations (\ref{a1}), (\ref{boua1}) where all
quantities should be calculated with the effective metric
$\hat g_{\mu\nu}=e^{2\hat\rho}g_{\mu\nu}$:
\begin{eqnarray}
&&\sqrt{\hat g} \hat E=\sqrt{g}(\phi_{,\mu\mu}-\phi_{,\mu}\phi_{,\mu})\,,
\qquad \sqrt{\hat g} \hat R =-2\sqrt{g}\partial^2\hat\rho \,,\nonumber \\
&&\hat k\sqrt{\hat h}=\sqrt{h} (k-\partial_N \hat\rho )\,.
\label{hats}
\end{eqnarray}
All indices in (\ref{hats}) are contracted with $g_{\mu\nu}$.
For Neumann boundary conditions we choose ${\cal S}=0$.
We obtain
\begin{eqnarray}
\delta W^{\mbox{\scriptsize ren}}(\Delta^{[\hat \rho]}_+)&=&
-\frac 1{24\pi}\left[ \int_M d^2x\sqrt{g} \delta\hat\rho
(-2\partial^2\hat\rho +6(\partial^2\phi )-6(\partial\phi )^2 )
\right. \nonumber \\
&&+\left. \int_{\partial M} dy\sqrt{h}
(2\delta\hat\rho (k-\partial_N\hat\rho )\mp 3\partial_N
(\delta\hat\rho ))\right]   \nonumber \\
&&+a_1^{(0)}(\delta\hat\rho ,\Delta^{[\hat \rho]}_+)\,.
\label{1525}
\end{eqnarray}
The $\mp$ sign in the surface integral corresponds to
Dirichlet ($-$) or Neumann ($+$) boundary conditions.
The last term in (\ref{1525}) describes the contribution of the
zero modes. It is present for Neumann boundary conditions
only and is given by (\ref{a1zero}) with $\phi\to -\phi$
and $\sqrt{g}\to e^{2\hat\rho }\sqrt{g}$. From this equation
we easily obtain
\begin{eqnarray}
&&W^{\mbox{\scriptsize ren}}(\Delta^{[\hat \rho]}_+)-
W^{\mbox{\scriptsize ren}}(\Delta_+) =
-\frac 1{24\pi}\left\{ \int_M d^2x\sqrt{g}
[(\partial\hat\rho )^2 +6\hat\rho (\partial^2\phi -(\partial\phi )^2)]
\right.\nonumber \\
&&\qquad\qquad +\left. \int_{\partial M} dy\sqrt{h}
(2\hat\rho\hat{k} \mp 3\partial_N \hat\rho )\right\} +
 \frac 14 (1\mp 1) \log 
\frac{\int_M d^2x e^{2(\hat{\rho} -\phi )}}{\int_M d^2x 
e^{ -2\phi }} \,.
\label{1620}
\end{eqnarray}
Here again the upper sign corresponds to Dirichlet and the lower
to Neumann boundary conditions. The action $W^{\mbox{\scriptsize scale}}$
in (\ref{scaling}) is given by the right hand side of the same equation
(\ref{1620}) with $\hat \rho =\rho -\varphi +\phi$. 
Now it is clear which modifications
should be made in (\ref{dWbou}) for arbitrary curved 2D manifolds 
(for a non-zero conformal factor
$\hat\rho$):
\begin{eqnarray}
W^{\mbox{\scriptsize ren}}(\Delta^{[\hat \rho]}_+)-
W^{\mbox{\scriptsize ren}}(\Delta^{[\hat \rho]}_-)&=&
\frac 1{4\pi} \int_M d^2x \sqrt{\hat g} \hat R \phi 
+\frac 1{2\pi }
\int_{\partial M} dy\sqrt{\hat h}  \hat k\phi \nonumber \\
&&-\frac 12 \log \left( \int d^2z
\sqrt{\hat g}e^{2\phi} \right) +W^{[0]} \,. \label{dWbour}
\end{eqnarray}
It is interesting to note that in the volume and surface integrals
the constant mode of the dilaton again couples to the Euler
characteristic.

\section{Dilaton Maxwell theory in 4D}
Consider the dilaton-Maxwell system in flat four dimensional
Euclidean space, 
\begin{equation}
S=\frac 14 \int d^4 x \, e^{-2\phi}
{\cal F}_{\mu\nu}{\cal F}^{\mu\nu} \,, \label{actdM}
\end{equation}
where ${\cal F}$ is an abelian field strength,
${\cal F}_{\mu\nu}=\partial_\mu a_\nu -\partial_\mu a_\nu$.
Such an interaction appears in extended supergravities \cite{Cremmer78},
string theory, and after a reduction from higher dimensions.
Mechanisms for the generation of the action (\ref{actdM}) and its
classical properties were considered e.g. in
\cite{Gibbons82,Maeda86,Frolov87,Gibbons88,Garfinkle91,Horowitz:1991cd}
(see also \cite{no98,odi98,io98} for
some quantum calculations).

If the action (\ref{actdM}) is obtained by a reduction from higher
dimensions, the inner products on the spaces of vector and scalar
(ghost) fields should also undergo the reduction and should thus
contain the dilaton field
\begin{eqnarray}
<a,a'>&=&\int d^4x\, e^{-2\phi} a_\mu {a_\mu}' \,,\nonumber \\
<\sigma ,\sigma'>&=&\int d^4x\, e^{-2\phi}\sigma \sigma' \,.
\label{measures}
\end{eqnarray}
It is convenient to choose the gauge condition as
\begin{equation}
(\partial_\mu -\phi_{,\mu} )e^{-\phi}a_\mu =0 \,.
\label{gaucond}
\end{equation}
In this gauge the action for $a_\mu$ and the Faddeev-Popov
ghosts $\sigma$ is
\begin{eqnarray}
&&S=\frac 12 \int d^4x \left[ a_\nu e^{-\phi} \left[
-(\partial_\mu -\phi_{,\mu })(\partial_\mu +\phi_{,\mu})
\delta_{\nu\rho} +2\phi_{,\nu\rho} \right) e^{-\phi} a_\rho
\right. \nonumber \\
&&\qquad\qquad -\left. \sigma e^{-\phi}(\partial_\mu
-\phi_{,\mu })(\partial_\mu +\phi_{,\mu})e^{-\phi}\sigma
\right] \,.\label{gfact}
\end{eqnarray}
Performing the functional integration over $a_\mu$ and $\sigma$
with the measure (\ref{measures}) we arrive at the
effective action
\begin{equation}
W=\frac 12 \log 
\frac{\det (\Delta_+\delta_{\mu\nu} +2\phi_{,\mu\nu})}{
\det (\Delta_+)} \,,\label{ea4}
\end{equation}
where $\Delta_+ =-(\partial_\mu -\phi_{,\mu })(\partial_\mu +\phi_{,\mu})$
and the vector determinant in (\ref{ea4}) is restricted to the fields
$e^{-\phi}a_\mu$ satisfying the gauge condition (\ref{gaucond}).

The aim of this section is to study transformations of the effective
action (\ref{ea4}) under change of the sign in front of the dilaton
field. We adapt to this problem the technique of the multidimensional
Darboux transform \cite{ABI}.

Let us introduce the multi-index Kroneker symbol,
\begin{equation}
\delta_{\mu_1\dots\mu_n}^{\nu_1\dots\nu_n}=\frac 1{(4-n)!}
\epsilon_{\mu_1\dots\mu_n\rho_{n+1}\dots\rho_4}
\epsilon^{\nu_1\dots\nu_n\rho_{n+1}\dots\rho_4} \,,
\label{Kron}
\end{equation}
and the ``supercharges''
\begin{equation}
Q^{(n)-}_{[\mu ][\nu ]}=
\delta_{\mu_1\dots\mu_n}^{\nu_1\dots\nu_{n-1}\rho} B_\rho
=\left( Q^{(n)+}_{[\nu ][\mu ]} \right)^\dag \,.
\label{Q+-}
\end{equation}
The operator $Q^{(n)-}$ maps $(n-1)$-forms to $n$-forms and
$Q^{(n)+}$ maps $n$ forms to $(n-1)$-forms.
With the help of these operators we define the matrix
``Hamiltonians''
\begin{eqnarray}
&&H^{(1)}_{\mu |\nu}=-B_\mu A_\nu \,,\qquad
\tilde H^{(1)}_{\mu |\nu}=\frac 12 Q^{(2)+}_{\mu |\rho_1\rho_2}
Q^{(2)-}_{\rho_1\rho_2 |\nu} \,,           \nonumber \\
&&H^{(2)}_{\mu_1\mu_2 |\nu_1\nu_2}=\frac 12 
Q^{(2)-}_{\mu_1\mu_2 |\rho}Q^{(2)+}_{\rho |\nu_1\nu_2 }\,,
\qquad \tilde H^{(2)}_{\mu_1\mu_2 |\nu_1\nu_2}= \frac 1{12}
Q^{(3)+}_{\mu_1\mu_2 |\rho_1\rho_2\rho_3}
Q^{(3)-}_{\rho_1\rho_2\rho_3 |\nu_1\nu_2 } \,,      \nonumber \\
&&H^{(3)}_{ \mu_1\mu_2\mu_3 |\nu_1\nu_2\nu_3}=\frac 1{12}
Q^{(3)-}_{\mu_1\mu_2\mu_3 |\rho_1\rho_2}
Q^{(3)+}_{\rho_1\rho_2 |\nu_1\nu_2\nu_3} \,.       \label{matrH}
\end{eqnarray}
These operators replace $F_{\mu\nu}$ and $\tilde F_{\mu\nu}$
of Sec. 2.
The operator $\tilde H^{(3)}$ is constructed in a similar
way. $H^{(n)}$ and $\tilde H^{(n)}$ act on the space ${\cal V}_n$
of the $n$-forms.
We observe that
\begin{equation}
Q^{(n)-}Q^{(n-1)-}=Q^{(n-1)+}Q^{(n)+}=0 \,.\label{QQ0}
\end{equation}
Thus $Q^-$ and $Q^+$ have the properties of the $d$ and $\delta$
operators of an elliptic complex.
There is an orthogonal decomposition ${\cal V}_n=
{\cal V}_n^+\oplus{\cal V}_n^-$ such that ${\cal V}_n^+=
Q^{(n-1)+}{\cal V}_{n+1}$ and ${\cal V}_n^-=
Q^{(n)-}{\cal V}_{n-1}$. The spaces ${\cal V}_n^+$ (resp. 
${\cal V}_n^-$) are spanned by the zero modes of $H^{(n)}$
(resp. $\tilde H^{(n)}$). If $\phi\to const.$ as $|x|\to\infty$
there are no more zero modes. In the construction of the
spectral zeta functions below we always assume that $H^{(n)}$
and $\tilde H^{(n)}$ are restricted to ${\cal V}_n^-$ and
${\cal V}_n^+$ respectively. 

The operators
\begin{equation}
{\cal H}^{(n)}=H^{(n)}+\tilde H^{(n)} \label{calH}
\end{equation}
are elliptic on the space of $n$-forms.
The effective action (\ref{ea4}) can be represented as
\begin{equation}
W=\frac 12 \log \frac {\det \tilde H^{(1)}}{\det \Delta_+} \,.
\label{eff4D}
\end{equation}
By acting as in Sec.\ref{2Dsec}, we write
\begin{equation}
\frac 12 \delta \zeta^+(0)'=-\zeta_0^+ (\delta\phi )
+\zeta_0^{H^{(1)}} (\delta\phi ) \,. \label{1526}
\end{equation}
The second term on the right hand side of (\ref{1526})
can be expressed as
\begin{equation}
\zeta_0^{H^{(1)}} (\delta\phi )=
\zeta_0^{{\cal H}^{(1)}} (\delta\phi )-
\zeta_0^{\tilde H^{(1)}} (\delta\phi ) \,. \label{1534}
\end{equation}
For the un-smeared $\zeta$-functions we have
\begin{equation}
\zeta^{\tilde H^{(1)}}(s)=\zeta^{{\cal H}^{(1)}} (s)
-\zeta^{H^{(1)}}(s)=
\zeta^{{\cal H}^{(1)}} (s)-\zeta^+ (s) \,.
\label{1542}
\end{equation}
The $\zeta$ functions on the right hand side of (\ref{1542})
correspond to elliptic operators of the Laplace type.
They are regular at $s=0$. Therefore, 
$\zeta^{\tilde H^{(1)}}(0)'$ is well defined, and
\begin{equation}
\zeta_0^{\tilde H^{(1)}} (\delta\phi )=
-\frac 12 \delta \zeta^{\tilde H^{(1)}}(0)' +
\zeta_0^{H^{(2)}} (\delta\phi ) \,.         \label{1554}
\end{equation}
Repeating these steps once again we obtain
\begin{eqnarray}
&&\zeta_0^{H^{(2)}} (\delta\phi )=
\zeta_0^{{\cal H}^{(2)}} (\delta\phi )-
\zeta_0^{\tilde H^{(2)}} (\delta\phi ) \,,     \nonumber \\
&&\zeta_0^{\tilde H^{(2)}} (\delta\phi )=
-\frac 12 \delta \zeta^{\tilde H^{(2)}}(0)' +
\zeta_0^{H^{(3)}} (\delta\phi )  \,.
\label{1556}
\end{eqnarray}
It is convenient to rewrite $H^{(3)}$ in the dual representation:
\begin{equation}
H^{(3)}_{\mu |\nu}=\frac 16 \epsilon^{\mu\mu_1\mu_2\mu_3}
\epsilon^{\nu\nu_1\nu_2\nu_3}
H^{(3)}_{\mu_1\mu_2\mu_3 |\nu_1\nu_2\nu_3}
=-\delta_{\mu\nu} B_\rho A_\rho +B_\nu A_\mu \,. \label{H3dual} 
\end{equation}
We also introduce  $\tilde H^{(3)}$ and ${\cal H}^{(3)}$
in the dual representation:
\begin{equation}
\tilde H^{(3)}_{\mu |\nu}=-A_\mu B_\nu \,,\qquad\qquad
{\cal H}^{(3)}_{\mu |\nu}=\delta_{\mu\nu}\Delta_- -2\phi_{,\mu\nu} \,.
\label{allH3}
\end{equation}
Since in the zeta functions in (\ref{1556}) all vector indices
are contracted we can evaluate them in the dual representation
as well:
\begin{equation}
\zeta_0^{H^{(3)}} (\delta\phi )=
\zeta_0^{{\cal H}^{(3)}} (\delta\phi )-
\zeta_0^{\tilde H^{(3)}} (\delta\phi )  \,.
\label{1731}
\end{equation}
Next we note that the operators $H^{(3)}$, ${\cal H}^{(3)}$ and
$\tilde H^{(3)}$ can be obtained from $H^{(1)}$, ${\cal H}^{(1)}$ and
$\tilde H^{(1)}$ by changing the sign in front of the dilaton
field $\phi$. Therefore,
\begin{equation}
\zeta_0^{\tilde H^{(3)}} (\delta\phi ) =-\frac 12 \delta \zeta^-(0)'
+\zeta^-_0(\delta\phi ) \,.\label{1736}
\end{equation}
Combining (\ref{1526}), (\ref{1534}), (\ref{1542}), (\ref{1554}),
(\ref{1556}), (\ref{1731}) and (\ref{1736}) we obtain
\begin{eqnarray}
&&\frac 12 \left( -\delta\zeta^+(0)' +\delta\zeta^{\tilde H^{(1)}}(0)'
-\delta\zeta^{\tilde H^{(2)}}(0)' + \delta\zeta^-(0)' \right)
\nonumber \\
&&\quad =\zeta_0^+ (\delta\phi ) -
\zeta_0^{{\cal H}^{(1)}} +\zeta_0^{{\cal H}^{(2)}}
-\zeta_0^{{\cal H}^{(3)}} +\zeta_0^- (\delta\phi ) \,.
\label{dzzeta}
\end{eqnarray}
The second line of (\ref{dzzeta}) contains elliptic operators
only. For them $\zeta_0^H (\delta\phi )=a_2(\delta\phi ,H)$. 
Consequently the second line of (\ref{dzzeta}) vanishes
by the Gauss-Bonnet theorem \cite{Gilkey95}. One can also check
this by direct calculations. Therefore,
\begin{equation}
W^{\mbox{\scriptsize ren}}[\phi ]=
W^{\mbox{\scriptsize ren}} [-\phi ] \,.
\label{symWren4}
\end{equation}

\section{Conclusions}
In this paper we have found several remarkable duality relations
between functional determinants. We have shown how
the effective action of quantum fields interacting with the 
(matrix-valued) scalar background field transforms under
inversion of the background field. For flat manifolds without boundaries
the effective action is proved to be invariant under this inversion.
This property has been explicitly demonstrated for two-dimensional
models and for the dilaton-Maxwell theory in four dimensions.
The presence of boundaries leads to 
additional terms that can also be calculated exactly (see
Eq.(\ref{dWbou})). We have studied in detail the 2D matrix
theory, the 4D dilaton-Maxwell system, and the 2D dilaton theory on
a manifold with boundary. We have considered the simplest topologies
of the field configurations only. Less trivial topologies can be
included in our approach by taking into account the zero-mode
contributions.
The application of our results and methods
include duality properties of strings, quantum black holes, 
and quantum systems obtained after reduction from higher
dimensions,
to mention a few. 
We are sure that this list can be considerably extended.

\section*{Acknowledgements}
We would like to thank M. Bordag, S. Solodukhin, and A. Wipf
for very useful discussions of related questions.
Research of one of the authors (DV) has been supported by
the Deutsche Forschungsgemeinschaft, grant Bo 1112/11-1,
the Alexander von Humboldt foundation and 
the FWF (Fonds
zur F\"{o}rderung der wissenschaftlichen Forschung) 
project P-12.815-TPH.
AZ is grateful to DAAD (Deutscher Akademischer Austauschdienst)
for its financial support
and Institute of Theoretical Physics at Friedrich
Schiller University (Jena) for hospitality.  
AZ is also grateful to the Killam trust for its support.

\end{document}